\documentclass[iop, apjl, twocolappendix]{emulateapj}
\usepackage{graphicx}
\usepackage{bm}
\usepackage{amsmath,amssymb}
\usepackage{natbib}
\usepackage{float}
\usepackage{subfigure}
\usepackage{units}
\usepackage{hyperref}
\usepackage{color}
\usepackage{pifont}
\usepackage{float}
\usepackage[normalem]{ulem}
\usepackage{totcount}

\newtotcounter{citnum} 
\def\oldbibitem{} \let\oldbibitem=\bibitem
\def\bibitem{\stepcounter{citnum}\oldbibitem}

\newcommand{\bilby}{{\sc Bilby}}

\newcommand{\pymultinest}{{\sc pyMultinest}}
\newcommand{\afterglowpy}{{\sc afterglowpy}}
\newcommand{\ciao}{{\sc ciao}}

\newcommand{\thetaobs}{10^{\circ}\pm3^{\circ}}
\newcommand{\thetacore}{4.4^{\circ}\pm0.9^{\circ}}
\newcommand{\thetawing}{10^{\circ}\pm3^{\circ}}
\newcommand{\betaposterior}{4.7\pm 1.7}
\newcommand{\nism}{1.8^{+0.6}_{-0.8}}
\newcommand{\pposterior}{2.2\pm0.1} 
\newcommand{\epse}{-0.7\pm 0.3}
\newcommand{\epsb}{-0.8\pm 0.5}
\newcommand{\xin}{0.8\pm0.2}
\newcommand{\eiso}{52.2\pm0.5}
\newcommand{\tburst}{16784^{+9}_{-13}}
\newcommand{\cdf}{CDF-S~XT1}

\newcommand{\citeg}[1]{\citep[e.g.,][]{#1}}

\begin{document}
\title{\cdf: The off-axis afterglow of a neutron star merger at $z=2.23$}
\author{Nikhil Sarin\altaffilmark{1}, Gregory Ashton, Paul D. Lasky, Kendall Ackley, Yik-Lun Mong, Duncan K. Galloway}
\shortauthors{Sarin et al.}
\affil{School of Physics and Astronomy, Monash University, Vic 3800, Australia\\
OzGrav: The ARC Centre of Excellence for Gravitational-wave Discovery, Clayton, Victoria 3800, Australia\\
Royal Holloway University of London, Egham, Surrey, TW20 0EX, U.K.}
\altaffiltext{1}{nikhil.sarin@monash.edu}

\begin{abstract}
CDF-S XT1 is a fast-rising non-thermal X-ray transient detected by \textit{Chandra} in the Deep-Field South Survey. Although various hypotheses have been suggested, the origin of this transient remains unclear. Here, we show that the observations of CDF-S XT1 are well explained as the X-ray afterglow produced by a relativistic structured jet viewed off-axis. We measure properties of the jet, showing that they are similar to those of GRB170817A, albeit at cosmological distances. We measure the observers viewing angle to be $\theta_{\textrm{obs}} = \thetaobs$ and the core of the ultra-relativistic jet to be $\theta_{\textrm{core}} = \thetacore$, where the uncertainties are the $68\%$ credible interval. The inferred properties and host galaxy combined with Hubble, radio, and optical non detections favour the hypothesis that CDF-S XT1 is the off-axis afterglow of a binary neutron star merger. We find that other previously suggested hypotheses are unable to explain all properties of CDF-S XT1. At a redshift of $z=2.23$, this is potentially the most distant observed neutron star merger to date and the first orphan afterglow of a short gamma-ray burst. We discuss the implications of a binary neutron star merger at such a high redshift for the star-formation rate in the early Universe, the nucleosynthesis of heavy elements, and the prospect of identifying other off-axis afterglows. 
\end{abstract}
\keywords{gamma-ray burst: general, X-rays: burst}

\section{Introduction}\label{sec:intro}
Gamma-ray bursts are highly energetic explosions caused by either the collapse of a massive star or the merger of a neutron star binary. While the mechanism behind the prompt gamma-ray emission is still uncertain~\citeg{peer15}, the weaker broadband afterglow emission observed extensively in X-rays, optical and radio has been studied in significant detail. This broadband afterglow is widely believed to be synchrotron radiation produced from the interaction of a relativistic jet with the surrounding interstellar medium~\citeg{sari99}.

It has long been accepted that gamma-ray bursts are collimated into an ultra-relativistic jet~\citeg{kumar15}, a point spectacularly confirmed by very-long baseline interferometry of GRB170817A~\citeg{mooley18_superluminal}, the gamma-ray burst that accompanied the first gravitational-wave observation of a binary neutron star merger~\citep{abbott17_gw170817_gwgrb,abbott17_gw170817_multimessenger}.
In general, as the jet ploughs into the surrounding interstellar medium, the initial high velocity of the material implies the emission is relativistically beamed and can only be seen for an observer close to the emission axis. 
As that material slows, the beaming cone broadens, and off-axis observers begin to see the burst afterglow.
This relativistic beaming necessarily implies the existence of \textit{orphan} afterglows, where the observer misses the early prompt emission but sees the afterglow as the jet broadens.

Orphan afterglows are notoriously difficult to find. Despite numerous searches in multiple diverse data sets, there are only three putative long gamma-ray burst orphan afterglow candidates, one observed in radio~\citep{marcote19} and two in optical~\citep{ho20}. In 2014, the Chandra Deep-Field South Survey detected a fast-rising transient known as \cdf{}~\citep{bauer17}. \cdf{} is associated with a host galaxy in the CANDELS survey~\citep{candels} with a photometric redshift of $z=2.23^{+0.98}_{-1.84}$ ($2\sigma$). With a non-thermal spectrum, a photon index $\gamma \approx 1.43$~\citep{bauer17} (i.e., a declining spectrum), and non-detections in optical and radio, the transient properties are unlike many other high-energy transients~\citep{bauer17}. Since its discovery, various hypotheses have been suggested to explain the observations and properties of \cdf{}. Hypotheses such as a supernova-shock breakout~\citep{bauer17,alp20}, a tidal disruption event of a white dwarf with an intermediate-mass black hole~\citep{bauer17, peng19}, and the trapped emission from a millisecond magnetar~\citep{sun19} have all been explored previously to varying success.

In this Letter, we show that \cdf{} observations can instead be well interpreted as the X-ray afterglow produced by a relativistic jet viewed off-axis.
In Sec.~\ref{sec:methods} we describe our new method for fitting the raw photon count data with detailed structured jet models. In Sec.~\ref{sec:offaxis} we discuss the results and present evidence for the observations being explained as the X-ray afterglow of a gamma-ray burst viewed off axis. In Sec.~\ref{sec:short} we utilise the multi-wavelength non detections and the spectrum to build evidence for \cdf{} being the X-ray afterglow of a short gamma-ray burst produced by the merger of a neutron star binary. In Sec.~\ref{sec:alternatives} we discuss the weaknesses of the alternative interpretations. We conclude and discuss the implications of our results in Sec.~\ref{sec:implications}. 
\section{Method}\label{sec:methods}
\cdf{} was observed on October 1, 2014. The first photon arrived $\unit[1.68\times10^4]{s}$ into the observation period~\citep{bauer17}. 
During the subsequent $\sim \unit[100]{s}$, the photon count rate sharply increased, before decreasing over the next $\sim \unit[10^5]{s}$.
We develop a new method for analysing the data of \cdf{} to fit the data with various models and infer parameters of the burst including the time of the prompt emission, density of the interstellar medium, the opening angle and energetics of the relativistic jet, and the observer viewing angle. We use individual photon arrival times from the source region of \cdf{} as defined in~\cite{bauer17} with a $0.26"$ spatial extraction region.
We use the Chandra Interactive Analysis of Observations~(\ciao)~\citep{ciao} software package to extract $115$ source photons in the $\unit[0.3-7]{keV}$ energy range. Given the location of the transient and the point-spread function, this observation is unlikely to suffer from pileup (two incident photons count as one, or get rejected)~\citep{bauer17}. There are no other photons in this region across the entire observation.

The first photon has been the source of debate, with~\cite{sun19} considering that photon to be consistent with the background. 
We analyse the entire $\approx\unit[5\times10^4]{s}$ observation epoch to determine whether this photon is consistent with the background. We calculate the background rate from an off-source region to be $3\times10^{-5} ~\unit{cts~s^{-1}}$. Given this rate, the probability that one background photon arrives within $\sim 50$\unit{s} of all the other photons is $1\times10^{-4}$, suggesting that it is unlikely this photon is consistent with the background. Conversely, given our fitting allows us to estimate the source rate, the probability that this photon is from the source is $\gtrsim 74$\%.

We use the raw photon time of arrivals from \textit{Chandra} that are binned with a constant bin size of $\Delta T = \unit[3.2]{s}$ across the entire observation period. We perform Bayesian inference using a Poisson likelihood on this binned data. Our likelihood is,
\begin{equation}\label{eq:likelihood}
\mathcal{L}(\vec{d}|\vec{\theta}) = \prod_{i} \frac{r_{i}\Delta T e^{-r\Delta T}}{k_{i}!},
\end{equation}
where $r_{i}$ is the rate of photons in the $i^{\rm th}$ bin, $k_{i}$ is the number of photons in that bin, and $\Delta T$ is the bin size.
We calculate the rate in the $i^{\rm th}$ bin using,
\begin{equation}\label{eq:rate}
r_{i} = \frac{\sigma_{i}\int_{\unit[0.3]{keV}}^{\unit[7.0]{keV}}F(t_{i}, \vec{\theta}, \nu) d\nu}{E_{i}} + \Lambda_{\rm{background}},
\end{equation}
where $\sigma_{i}$ is the effective area of the detector, $E_{i}$ is the averaged energy of the photons in the given area, $F(t_{i}, \vec{\theta}, \nu)$ is the flux density of our model at time $t_{i}$ at a given frequency $\nu$ with associated vector of parameters $\vec{\theta}$, and $\Lambda_{\rm{background}}$ is the background rate. 
We note that this expression is an approximation and calculation of the true rate requires modelling the detector response and point-spread function. However, the uncertainties caused by this approximation are small compared to model uncertainties.
We calculate the flux density using the smooth power-law structured jet model in \afterglowpy{}~\citep{afterglowpy} with inverse Compton emission and jet spreading. We note that we also perform inference without these effects and our conclusions do not change. The smooth power-law structured jet model defines the energy distribution of the jet as, 
\begin{equation}\label{eq:jetstructure}
E(\theta_{\mathrm{obs}})=E_{0}\left(1+\frac{\theta_{\mathrm{obs}}^{2}}{\beta \theta_{\mathrm{core}}^{2}}\right)^{-\beta / 2}.
\end{equation}
Here $\beta$, is the exponent dictating the slope of the power-law jet structure. The Lorentz factor of the jet follows $E_{\theta_{\textrm{obs}}}^{1/2}$, with an initial Lorentz factor $\Gamma_0$. There is an additional parameter, $\theta_{\rm{wing}}$ which is a truncation angle outside of which the energy is initially zero. 
This structured jet interacts with the surrounding interstellar medium accelerating a fraction of electrons, $\xi_{n}$, with a fraction of the total thermal energy, $\epsilon_{e}$ and fraction of the thermal energy in the magnetic field, $\epsilon_b$. The synchrotron radiation produced by these electrons is responsible for the emission we observe.
We set the flux before the onset of the burst $T_{\rm{start}}$ to zero.
We sample over the redshift by putting a uniform prior on $z$ between $0.39-3.21$ corresponding to the 95\% credible interval from the host galaxy photometric redshift. We use Planck-15 cosmology to convert the redshift into a luminosity distance to the source~\citep{planck15}. 

To reduce the computational cost, we use a larger binning size of $\Delta T = \unit[64]{s}$ compared to the raw data. We verify there is no systematic bias introduced by this larger bin size by repeating the calculation with different bin sizes with \afterglowpy{}. We infer consistent parameters across $\Delta T = \unit[64, 128, 256]{s}$. We note that bin sizes larger than $256\unit{s}$ destroy the structure of the data and change the results.  
To ensure there is no bias introduced by not using the raw data, we also train a neural network algorithm to compute the flux density, calibrating it to the output of \afterglowpy{}. 
This trained model can be evaluated approximately three orders of magnitude faster than \afterglowpy{} making the analysis tractable with the raw data, but it introduces a systematic uncertainty due to the nature of machine learning. We infer consistent posteriors (at $1-\sigma$) with the neural network model on the raw data binned at $\Delta T = \unit[3.2]{s}$ and the larger bin sizes with \afterglowpy{}.
We have also verified that our machine learning model is accurate by performing tests on untrained data, simulating signals generated with \afterglowpy{} and recovering them with consistent parameters with the trained neural network model using the infrastructure described here.  
We will present details of the neural network model in future work.
\section{An off-axis gamma-ray burst}\label{sec:offaxis}
Several properties of \cdf{} are consistent with a gamma-ray burst afterglow, including the event rate~\citep{bauer17, sun19}. 
Firstly, the energetics and high redshift ($z=2.23^{+0.98}_{-1.84}$) demands a highly energetic transient, with upwards of $\unit[10^{50}]{erg}$ of energy. The non-thermal spectrum immediately rules out any thermal transient such as a supernova, while the fast rise and slow decay rule out a persistent non-thermal source. These properties, in particular the rise and slow decay are indicative of an off-axis gamma-ray burst afterglows~\citep{granot02}. Moreover, the declining spectrum and rising lightcurve is likely due to misaligned evolution i.e., $\theta_{\rm{obs}} > \theta_{\rm{core}}$~\citeg{afterglowpy}. 

To investigate the above interpretation in detail, we first consider the X-ray data. We apply our method described in Sec.~\ref{sec:methods} and fit the X-ray data of \cdf{}. We infer the parameters of the system using \bilby~\citep{bilby} and the \pymultinest{} sampler~\citep{multinest}. In Tab.~\ref{table:priors} we list the full set of parameters $\vec{\theta}$, their descriptions, their associated priors used in the analysis, and their posteriors with $68\%$ posterior credible interval. 

\begin{table*}
\centering
 \caption{Parameters associated with the smooth power-law structured jet model along with a brief description, the prior used in our analysis and the posterior. We note that the posterior values quoted here are from analysis with the neural network model run on data binned at $\Delta T = 3.2\unit{s}$.}
 \label{table:priors}
 \begin{tabular}{lccc}
  \hline
  Parameter [unit] & Description & Prior & Posterior\\
  \hline
$T_{\textrm{start}}$ [s] & burst start time into the observation & $ \textrm{Uniform}[10,16800]$ & $\tburst$ \\
$z$ & redshift & $\textrm{Uniform}[0.39,3.21]$ & $1.7^{+0.8}_{-0.7}$\\
$\Gamma_0$ & initial Lorentz factor & $\textrm{Uniform}[1,1000]$ & $420^{+300}_{-240}$\\
$\Lambda_{\rm{background}}$ [counts/s] & background rate & $\log\textrm{Uniform}[10^{-6},10^{-4}]$ & $10^{-4}\pm10^{-5}$\\\
$\theta_{\textrm{obs}}$ [$^\circ$] & observers viewing angle & $\textrm{Cosine}[0,0.7]$ & $\thetaobs$ \\
$\log_{10} (E_{\textrm{iso}}/\rm{erg}) $ & isotropic-equivalent energy & $\textrm{Uniform}[10^{44},10^{54}]$ & $\eiso$\\
$\theta_{\textrm{core}}$ [$^\circ$] & half-width of jet core& $\textrm{Uniform}[0.01,0.1]$ & $\thetacore$\\
$\theta_{\textrm{wing}}$ [$^\circ$] & wing truncation angle of the jet &$\textrm{Uniform}[0.01,0.7]$ &$\thetawing$\\
$\beta$ & power for power-law structure & $\textrm{Uniform}[0.5,10]$&$\betaposterior$\\
$\log_{10} (n_{\textrm{ism}}/\unit{cm^{-3}})$ & number density of ISM & $\textrm{Uniform}[-5,2]$& $\nism$\\
$p$ & electron distribution power-law index & $\textrm{Uniform}[2,3]$& $\pposterior$\\
$\log_{10}\epsilon_{e}$ & thermal energy fraction in electrons& $\textrm{Uniform}[-5,0]$& $\epse$\\
$\log_{10}\epsilon_{b}$ & thermal energy fraction in magnetic field & $\textrm{Uniform}[-5,0]$& $\epsb$\\
$\xi_{N}$& fraction of accelerated electrons & $\textrm{Uniform}[0,1]$& $\xin$\\
  \hline
 \end{tabular}
\end{table*}

\begin{figure*}[ht!]
    \centering
    \includegraphics[width=0.7\textwidth]{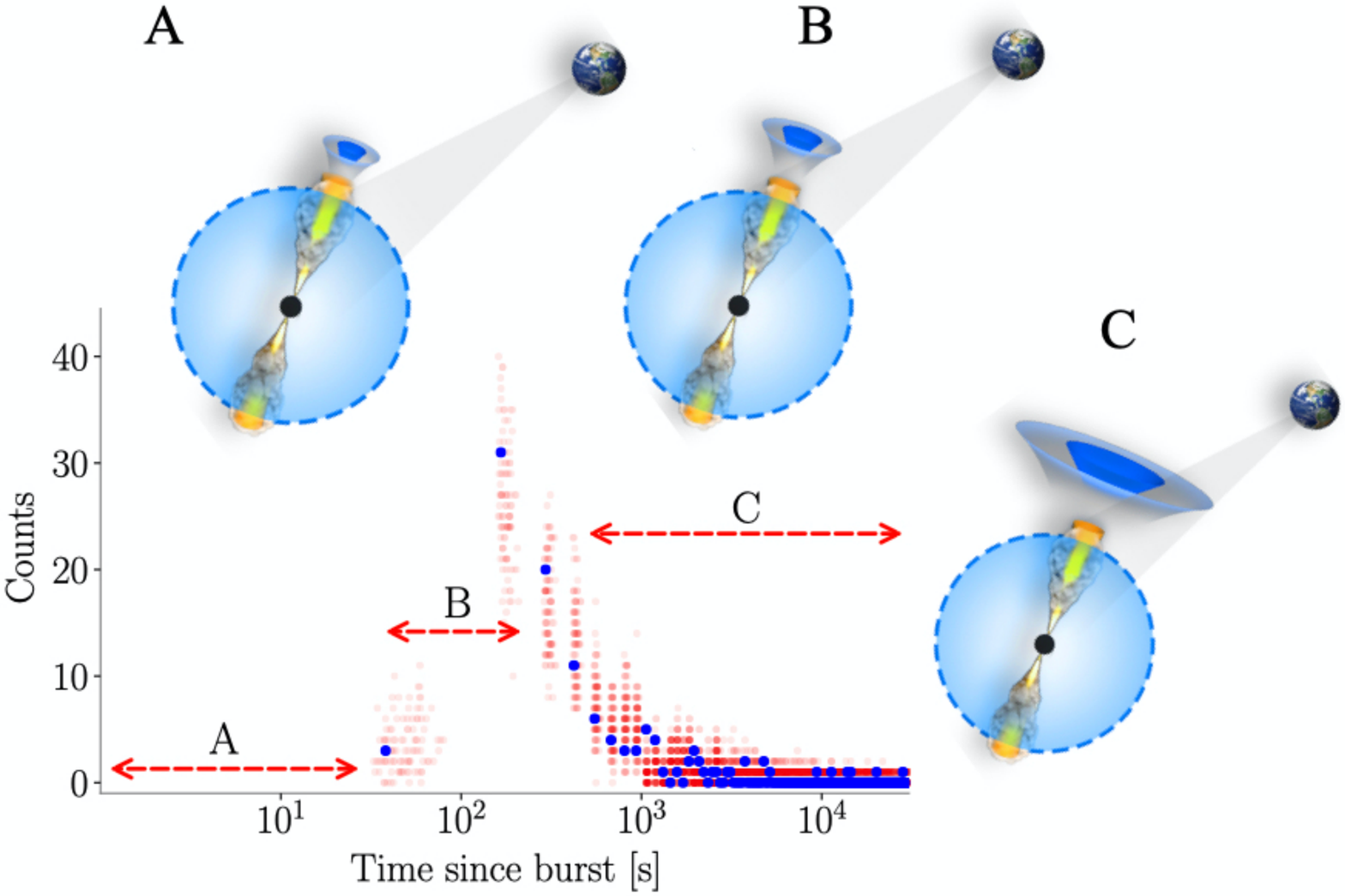}
    \caption{\textbf{\cdf{} data and model light curves}. X-ray counts as a function of time binned at $\Delta T = 128$s intervals in blue, and posterior predictions for the off-axis short gamma-ray burst model in red. The schematic illustrates the physical process responsible for the different observational epochs. In epoch A, the observer sees no emission as the ultra-relativistic jet (blue cone) and mildly relativistic wings (grey cone) are beamed away from Earth's line of sight. As the structured jet slows, its beaming cone broadens such that the mildly relativistic wings are now in Earth's line of sight. At this point, the observer on Earth starts to see X-ray photons from the source (epoch B). As the jet slows down further it continues to broaden at which point the jet core becomes visible to the observer on Earth and the photon rate peaks (epoch C). The subsequent gradual decay is a product of further slowing down of the jet and subsequent broadening of the jet.}
    \label{fig:lightcurveandcounts}
\end{figure*}

In Fig.~\ref{fig:lightcurveandcounts}, we plot the raw X-ray photon counts detected by Chandra in the $\unit[0.3-7.0]{keV}$ band. For visualisation purposes, we bin the photon arrival times with a bin size of \unit[128]{s}. Times in Fig.~\ref{fig:lightcurveandcounts} are referenced to our estimated burst start time.
In red, we show model-predicted counts from 100 random draws of the posterior distribution. In the same Figure, we also show a schematic view of the physics creating the various features of the light curve.  In particular, the jet slowing and subsequent broadening of the beaming cone give rise to the sharp increase in photon count seen approximately \unit[100]{s} after the burst. 

We measure the observer angle to be $\theta_{\rm obs} = \thetaobs$ and the core of the jet to be $\theta_{\rm core} = \thetacore$. This implies that \cdf{} was observed from outside the jet-core naturally explaining the lack of prompt gamma-ray emission due to relativistic beaming. 

We derive the start time of the burst to be $T_{\rm{start}} = \unit[\tburst]{s}$ into the observation which implies that the afterglow peaks approximately $\unit[150]{s}$ after the predicted time of the prompt emission.
Within our off-axis interpretation, this peak timescale $t_{\rm{peak}}$ is dictated by the timescale for the edge of the ultra-relativistic core to start becoming visible to the off-axis observer i.e., for $\Gamma(\theta_{\rm{obs}} - \theta_{\rm{wing}}) \sim 1$. A back-of-the-envelope estimate for this peak timescale is~\citeg{nakarpiran20}
\begin{equation*}
t_{p} \approx \unit[130]{d} \left(\frac{E}{\unit[10^{51}]{erg}} \frac{\unit[10^{-3}]{cm^{-3}}}{n_{\rm{ism}}}\right)^{1/3} \left(\frac{\theta_{\rm{obs}} - \theta_{\rm{wing}}}{15^{\circ}}\right)^2.
\end{equation*}
From our estimated parameters, this gives $t_{p} \sim \unit[700]{s}$, which is comparable to the observations. 

We note that this peak timescale is different to the peak timescale seen for GRB170817A, which came $\unit[160]{d}$ following the prompt. That peak time was the jet break time. For \cdf{} this can be estimated by~\citeg{granot17}
\begin{equation*}
t_{\rm{break}} = \unit[70]{d} E_{51}^{1/3} n_{0}^{-1/3}\theta_{\rm obs}^2.
\end{equation*}
From our estimated parameters this is $t_{\rm{break}} \sim \unit[4]{days}$,
which is significantly shorter than the $t_{\rm{break}}$ of GRB170817A. 
However, this discrepancy in timescales is not inconsistent and can be ascribed to the smaller viewing angle $\theta_{\rm{obs}}$ and the higher interstellar medium density $n_{\rm ism}$ for \cdf{}. 

Our inferred estimates for the isotropic energy is also consistent with a back-of-the-envelope estimate. The reported fluence for \cdf{} is $4.2^{+3.5}_{-0.2}\times 10^{-9} \unit{erg~cm^{-2}}$~\citep{bauer17}. At a redshift of $z=2.23$, this implies a jet with energy $\sim \unit[10^{50}]{erg}$, which is comparable to the estimated jet energy from our analysis.
Comparing the energetics \cdf{} to X-ray afterglows seen in other gamma-ray bursts, the total energy of \cdf{} would be lower by two orders of magnitude compared to the dimmest X-ray afterglows we see unless \cdf{} is at $z \gtrsim 2$. Although, since \cdf{} is observed off-axis, some difference in brightness is to be expected. 

Although Figure~\ref{fig:lightcurveandcounts} shows the fit on a $\unit[128]{s}$ binned timescale, we infer similar parameters across all bin sizes i.e., our conclusion is robust to the choice of binning. Similarly, even accounting for the redshift uncertainty, our results and conclusion are quantitatively similar: \cdf{} is consistent with being an orphan afterglow of an off-axis gamma-ray burst. We note that the posterior values mentioned above and those listed in Tab.~\ref{table:priors} are marginalised over the redshift uncertainty. 

To probe the nature of \cdf{}, the source region was searched in several other electromagnetic bands. In optical, the Very Large Telescope (VLT) observed the location of the transient $80$ minutes after the arrival of the first photon, and again $18$ days later, setting stringent $r$-band upper limits~\citep{triester14a, bauer17}.  
The Hubble Space Telescope (HST) searched the source region $111$ days after the first X-ray observations, setting the most sensitive limits for the source region in the $F110W$ filter~\citep{bauer17}. 
In radio, the Australian Telescope Compact Array (ATCA) observed the source region seven days after the first X-ray photons, setting limits at different radio frequencies~\citep{burlon14}. 
In gamma rays, neither {\it Swift} nor {\it Fermi} had coverage in the direction of \cdf{} for the few hours surrounding the event~\citep{bauer17}. 
None of these observations found a counterpart to \cdf{}. 
To correctly identify the nature of \cdf{}, we must be able to explain these non-detections. Gamma-ray burst afterglow models make reliable predictions about emission across all of these bands. Therefore these non-detections can be used to verify our proposed off-axis gamma-ray burst hypothesis.

\begin{figure*}[!ht]
    \centering
    \includegraphics[width=\textwidth]{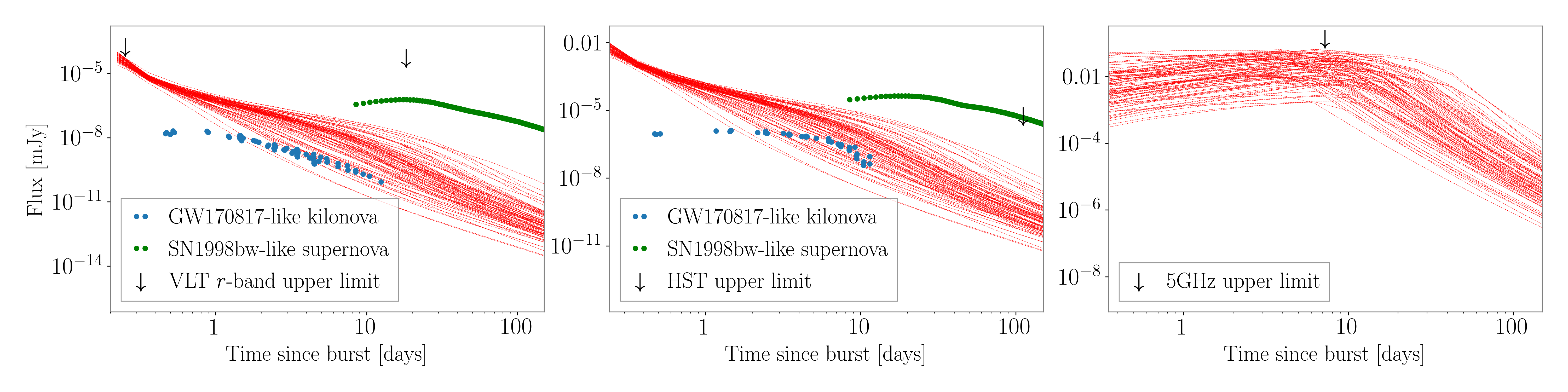}
    \caption{Optical and radio upper limits are above predictions from our model.  Flux density in optical ($\lambda=640$nm; left panel, and $\lambda=1179$nm; middle panel) and radio ($5$GHz; right panel), with arrows representing VLT, HST and ATCA upper limits, respectively~\citep{bauer17}.
    Each red curve is a predicted light curve, calculated by drawing parameters randomly from our posterior probability distribution.
    The arrows correspond to upper limits derived from VLT, HST and ATCA observations respectively~\citep{triester14a,burlon14}. 
    The blue points are data from the kilonova associated with GW170817 but scaled assuming a redshift of $z=2.23$. Likewise, the green points are the data for supernova SN1998bw but scaled to a redshift of $z=2.23$.  Our posterior predictive models are consistent with the interpretation of this event as a compact object coalescence, but not necessarily with that of a supernova.}
    \label{fig:optical}
\end{figure*}

We evaluate the predicted flux density light curves in optical at \unit[640]{nm} and \unit[1179]{nm}, and radio at \unit[5]{GHz}, corresponding to the VLT $r$-band, HST $F110W$-band, and ATCA observations, respectively.  We show these lightcurves, as well as the upper limits in optical and radio in Fig.~\ref{fig:optical} at a fixed redshift $z = 2.23$ corresponding to the median value of the host galaxy photometric redshift for easier comparison.
The left and middle panels show the optical light curves for $100$ different samples randomly drawn from our posterior distribution as predicted in the VLT $r$-band and the HST $F110W$-band, while the right panel shows the corresponding flux density in the radio band. The black arrows indicate the upper limits set by VLT, HST and ATCA in the left, middle and right panels, respectively.

We estimate the total extinction from its association with the neutral hydrogen column density, which we infer from the spectra. We infer a total extinction of $A_V = 5.96^{+6.98}_{-3.22}$ at a redshift $z = 2.23$, largely due to the strong redshift dependence of the inferred hydrogen column density~\citep{bauer17}. 
In Fig.~\ref{fig:optical},   
These observed upper limits are above the corresponding optical and radio light curve predictions, consistent with the non-detections in these electromagnetic bands. 
We note that in Fig.~\ref{fig:optical} we have used the median value of $A_V = 5.96$ as the total extinction in the $r$-band. The $F110W$-band and radio do not suffer such high extinction.
Given our uncertainty on the extinction, we estimate that there is a less than $10^{-4}$ \% probability that the optical afterglow in $r$-band would have been detectable. 

In gamma rays, neither {\it Swift} nor {\it Fermi} had coverage in the direction of \cdf{} for the few hours surrounding the event~\citep{bauer17}. 
Our inferred parameters from fitting the X-ray data and the large redshift implied from the identification of the host galaxy imply no prompt emission would have been detected from this system even with gamma-ray coverage. Furthermore, the inferred parameters can explain the non-detections in other electromagnetic bands. 
This analysis above suggests \cdf{} is potentially the first orphan afterglow detected in X rays and one of the most distant orphan afterglow candidates ever. 

\section{A compact object merger?}\label{sec:short}
Having established that \cdf{} is potentially an orphan X-ray afterglow of a gamma-ray burst we now turn to answer whether \cdf{} is an orphan afterglow of a long or short gamma-ray burst?  In other words, was this the orphan afterglow due to the collapse of a massive star, or the merger of a neutron star binary? Here, we investigate the various properties of \cdf{} and the multi-wavelength observations to answer this question. 

A short gamma-ray burst would likely be accompanied by an optical/infrared kilonova, such as AT2017gfo for GW170817~\citeg{abbott17_gw170817_multimessenger} and several other short gamma-ray bursts.
Similarly, a long gamma-ray burst should be accompanied by a core-collapse supernova, such as the case for GRB980425 accompanied by SN1998bw~\citeg{sn1998bw} and GRB111209A accompanied by SN2011kl~\citeg{grenier2015}.
In the left and middle panels of Fig.~\ref{fig:optical} we show the prediction for a GW170817-like kilonova (blue points) and a SN1998bw-like supernova (green points) if either had accompanied \cdf{} at $z=2.23$. We pick SN1998bw as it is one of the dimmest supernovae to accompany a long gamma-ray burst, and largely due to its proximity, the best observed.
In particular, we take data from AT2017gfo and scale it to a distance of $z=2.23$.  The VLT upper limits being above the blue points indicates that the kilonova associated with GW170817 would not have been seen by VLT had it been at a redshift of $z=2.23$. Moreover, that the red predictive curves are consistent with the GW170817-like kilonova indicate too that, if \cdf{} was a short gamma-ray burst with associated kilonova, one would not have expected to see it with VLT observations.

The green dots on the left and middle panels of Fig.~\ref{fig:optical} take the $r$ and $i$-band observations from SN1998bw, respectively, and scale them to a distance of $z=2.23$. We note that SN1998bw was not observed by HST and the $i$-band observations are likely a lower estimate of what SN1998bw would have looked like realistically with HST. 
SN1998bw was a relatively dim supernova compared to other supernovae that accompany long gamma-ray bursts~\citeg{sn1998bw}. That the green dots are close to the HST upper limit implies an SN1998bw-like supernova would have been marginally detectable.  Had a brighter supernova been associated with \cdf{}, it would have been potentially observed by the HST. This deep constraint from HST rules out most potential supernovae up to a redshift $z \sim 2$~~\citep{richardson14}. A similar point could be made for the VLT observations if this system does not suffer from severe extinction.
If a supernova accompanied \cdf{}, it would have had to be relatively dim to not be observed by HST. We note that there have been observations of long gamma-ray bursts without an accompanying supernova, however, these are likely due to observational selection effects~\citep{lyutikov13}.
This non-detection of a supernova adds weight to the hypothesis that the progenitor of \cdf{} was the merger of a neutron star binary rather than the collapse of a massive star. 

The star-formation rate ($1.15 \pm 0.04 ~M_{\odot}$\unit[yr]$^{-1}$) of the putative host galaxy CANDELS $28573$ and the off-set ($0.13"$) are also consistent with other short gamma-ray bursts~\citep{bauer17,sun19}. We note that the uncertainty in the location of the transient is smaller than the size of a typical galaxy, such that the effect of the incompleteness of galaxy catalogues does not significantly affect this association. Furthermore, given the implied high redshift, a requirement based on energetics, the effect of the supernova kick also does not add significant uncertainty to affect this association. We note that our inferred interstellar medium density suggests that the gamma-ray burst was not significantly offset from the host.
The mass of the host galaxy is on the smaller end for short gamma-ray bursts, while the star formation rate is comparable. However, the limited number of known short gamma-ray bursts beyond redshift $z\approx2$ implies this is not a statistically robust statement~\citep{sun19}. 

Our inferred interstellar-medium density $n_{\textrm{ism}}=\unit[\nism]{cm^{-3}}$ is more akin to long gamma-ray bursts than short~\citep{fong15}.  However, approximately 5 to 20\% of short gamma-ray bursts~\citep{fong15} have $n_{\rm ism}\gtrsim\unit[1]{cm^{-3}}$.
Moreover, systematic studies of interstellar-medium densities to date typically fix the energy fraction in the magnetic field and electrons, which systematically underestimates $n_{\rm ism}$. 
We do not fix these parameters but instead marginalise over this uncertainty.


Finally, we consider the implied spectrum of \cdf{}. Gamma-ray bursts show an empirical correlation between the isotropic gamma-ray energy and the rest-frame peak energy in gamma-rays. For long gamma-ray bursts this is known as the Amati correlation~\citep{amati06}, which differs from that measured for short gamma-ray bursts with observed redshifts. Therefore, identifying the consistency with the Amati relation offers another way to probe whether the gamma-ray burst is long or short~\citeg{amati06, minaev20}. 

\begin{figure}[!ht]
    \includegraphics[width=0.5\textwidth]{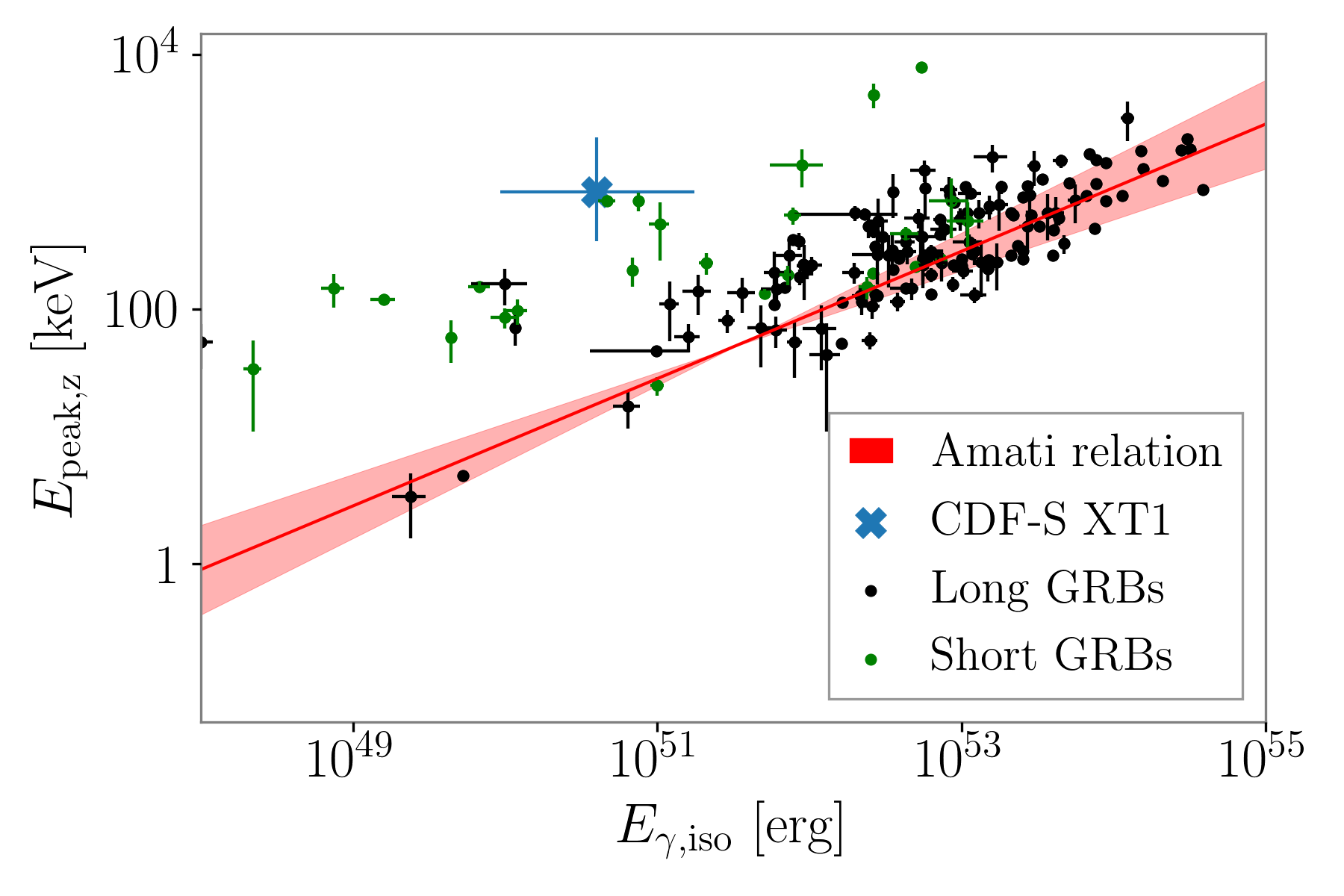}
    \caption{The observed isotropic gamma-ray energy and peak rest-frame energy of \cdf{} (blue) and a sample of observed long (black) and short (green) gamma-ray bursts~\citep{minaev20}. The empirical Amati relation and associated uncertainty is shown in red. The uncertainty on \cdf{} is obtained by propagating uncertainties in the spectral slope, the inferred kinetic energy and the distribution of gamma-ray radiative efficiencies~\citeg{fong15}.~\cdf{} is inconsistent with the Amati relation, while being consistent with observed short gamma-ray bursts, adding weight to our argument that \cdf{} is the X-ray afterglow of a binary neutron star merger. Note that the Amati relation is derived from a significantly larger sample of observed gamma-ray bursts than shown here~\citep{minaev20}}
    \label{fig:amati}
\end{figure}

We use our inferred isotropic energy from the afterglow and the observed distribution of prompt emission gamma-ray efficiencies~\citep{fong15} to determine the isotropic energy in gamma rays emitted from \cdf{}, finding $\log_{10} (E_{\gamma, \rm{iso}}/\unit{erg}) = 50\pm 0.6$. We use the relation between the X-ray photon index and rest-frame peak energy~\citep{virgili12} to infer the rest-frame peak energy, $E_{\rm{peak, z}} = 830^{+1200}_{-500}~\unit{keV}$. We emphasise that this analysis is agnostic to whether a gamma-ray burst is from the collapse of a massive star or the merger of a neutron star binary. 
In Fig.~\ref{fig:amati}, we show that the inferred parameters for \cdf{} are inconsistent with observed long gamma-ray bursts and the Amati relation. However, these parameters are consistent with several observed short gamma-ray bursts. This adds significant weight to the hypothesis that \cdf{} is the orphan afterglow of a short gamma-ray burst produced in the merger of a neutron star binary.

\section{Alternative interpretations}\label{sec:alternatives}
Three alternate explanations were offered for this transient in the discovery paper~\citep{bauer17} and explored briefly elsewhere~\citep{sun19, peng19, alp20}.  Here we critically assess each of these scenarios, providing a summary of the different scenarios and their failure to explain observations in Tab.~\ref{table:scenarios}.

\begin{table*}
\centering
 \caption{Class of systems and emission mechanisms and whether they can explain the different observational constraints on \cdf{}.} 
 \label{table:scenarios}
  \begin{tabular}{lccccc}
  \hline
System & Emission mechanism & Spectrum & Luminosity & Optical constraints & Host-galaxy properties \\
  \hline
Tidal disruption & Tidal disruption event & ? & x & \checkmark &\checkmark \\
 & Jetted tidal disruption event & ? & x & ? &\checkmark \\
  \hline
Collapse of massive star & Supernova-shock breakout & x & x & x & \checkmark\\
 & Long gamma-ray burst orphan afterglow & x & \checkmark & ? & \checkmark \\
\hline
Compact-object coalescence & Short gamma-ray burst orphan afterglow & \checkmark & \checkmark &\checkmark &\checkmark\\
  & Kilonova with energy injection & ? & \checkmark & ? & \checkmark\\
  \hline
 \end{tabular}
\end{table*}

It has been argued that \cdf{} could be a supernova-shock breakout from a core-collapse supernova~\citep{bauer17,alp20} in which the shock wave from the supernova blasts out of the progenitor star's surface. While the timescale and steep-rise/gradual decay of the light curve is consistent with such phenomena, the spectrum is not.
Supernova shock-breakouts are thermal transients and the radiation they produce is typically soft X rays or ultra-violet depending on the temperature of the system~\citeg{tominaga11}. 
For typical shock breakouts, the dominant emission is expected at $\unit[0.01-1]{keV}$ i.e., a thermal spectrum, inconsistent with the non-thermal spectrum of \cdf{}.
Furthermore, the peak luminosity from typical supernova-shock breakouts~\citep{tominaga11} of $\unit[10^{44}-10^{45}]{erg}$ is inconsistent with the inferred peak luminosity of \cdf{} at $z\gtrsim0.4$.
To explain the observations of \cdf{} at $z=2.23$, a supernova-shock breakout would have to be more energetic than the most energetic shock-breakout observed by two orders of magnitude.
Importantly, a supernova-shock breakout will also be accompanied by a supernova.
However, the upper limits set by HST observations confidently rule out a supernova at redshift $z\lesssim2$~~\citep{richardson14}.
The combination of these constraints suggests that the supernova-shock breakout scenario is a poor explanation of the observations: the supernova-shock breakout would have to be exceptionally bright but accompanied by a relativity dim supernova and explain the hard spectrum. While we cannot definitively rule out a supernova-shock breakout, the event would be like nothing else we have seen before. 

It has been suggested that \cdf{} could be caused by a tidal disruption event~\citep{bauer17,peng19}. The fast rise and hard X-ray flux of \cdf{} imply that the only viable part of the parameter space for a tidal disruption event that can explain the observations is a compact white dwarf around an intermediate-mass black hole of around $~10^{3}-10^{4} M_{\odot}$~\citep{bauer17,peng19}.
However, such a scenario is problematic because the Eddington limit for such a system is $2-3$ orders of magnitude below what is required given the associated redshift~\citep{bauer17}.
This discrepancy in energy could be explained by a strongly beamed, jetted tidal disruption event. One would expect such a scenario to also be observed in radio and optical as is the case for other jetted tidal disruptions~\citep{cenko11} which also have longer decay timescales. However, we do not observe radio and optical counterparts and a long decay timescale is inconsistent with the decay of \cdf{}. The only proposed way to generate sufficient magnetic fields to create a jetted tidal disruption event is by anchoring the magnetic field to a black hole~\citep{tchek14} or through a highly magnetised star from a recent merger~\citeg{mandellevin15}. In both cases, the timescale for the rise and decay of the light curve would be significantly longer than that observed in \cdf{} and inconsistent with the observations.

Finally, a more exotic scenario was developed to explain both \cdf{} and CDF-S XT2 in a unified framework as being the emission from a millisecond magnetar combined with the kilonova from the ejecta of a neutron star merger~\citep{sun19}. 
In this scenario, \cdf{} was a neutron star merger that produced a remnant neutron star that acted as a central engine injecting energy into the surrounding ejecta.
To explain the lack of detection in optical, such a system would need to be unusually dim given the additional energy injection, which is possible for comparatively lower ejecta masses than for the kilonova associated with GW170817.
However, the ejecta cannot be too small either, as such a system would then not be optically thick and not explain the rise seen in the light curve of \cdf{}.
It is important to note that the typical timescale for such a system to become optically thin is $\sim\unit[9]{hr}$~\citep{metzger14}, which would imply such a model is inconsistent with \cdf. However, there is substantial latitude in the timescale given the uncertain opacity of the ejecta.
Furthermore, the spectrum from such a model would be a combination of the non-thermal spectrum from the spin-down of the nascent neutron star and quasi-thermal spectrum from the ejecta, which is at odds with the non-thermal spectrum inferred for \cdf{}. Moreover, this model requires the formation of an infinitely stable neutron star 
We note that to explain the observations of \cdf{} with such a model, ~\cite{sun19} considered the first photon of the observation as being consistent with the background---in Sec.~\ref{sec:methods} we argue that the first photon was from \cdf{}, implying their fit to \cdf{} is inconsistent with the data.

Even with an afterglow interpretation, there are other ways to interpret the data. Within our interpretation, to explain the relatively fast rise and subsequent decay, the data requires the jet to be truncated near the observer. The angle at which the jet truncates is set by $\theta_{\rm{wing}}$, which we keep as a free parameter.
This implies that the jet energy drops sharply at an angle close to the observer. Such a sharp drop is 
implicitly built into other jet models such as the top-hat~\citeg{afterglowpy}.
We analyse the data by not allowing such a sharp drop in energy by fixing $\theta_{\rm{wing}} = f\times\theta_{\rm{core}}$ where $f$ is some factor such that the jet energy is $\approx 0$, i.e., we do not truncate the jet. Such a model is perhaps more plausible than one with a sharp drop, although without detailed hydrodynamical simulations or extensive observations of off-axis gamma-ray bursts one cannot know for certain what the true jet structure is. We note that past hydrodynamical simulations do produce sharp drops in jet energy~\citeg{aloy05}.

If instead, we demand a jet structure that does not truncate, we would need to explain the rise time with the pre-deceleration behaviour of an on-axis relativistic jet. 
Fixing $f=8$, and fitting to the data, we find that the data can be explained by having a weakly relativistic jet with Lorentz factor $\Gamma_0 = 46^{+30}_{-20}$ and the observer being on-axis. In such a scenario, the rise and decay can be explained by the deceleration of a low Lorentz factor ($\lesssim 100$) jet. Such a relativistic jet may not produce prompt gamma-rays due to the pair production threshold for gamma-ray production, also known as the compactness problem. However, the threshold for prompt gamma-ray emission is not well understood. Such gamma-ray bursts are also known as failed gamma-ray bursts or dirty fireballs~\citeg{huang02, rhoads03}. We note, however, that this solution is not preferred by the data, with a Bayesian odds, assuming both hypotheses are equally likely, of $\mathcal{O} = 3.3$ in favour of truncating the jet. That is, the solution suggesting the jet truncates near the observer is $\sim 3$ times more favourable than not. 
\section{Implications and Conclusions}\label{sec:implications}
Considering the limitations of the other three models posed in the literature (see Sec.~\ref{sec:alternatives} and Tab.~\ref{table:scenarios} for a summary) and the success of the orphan afterglow interpretation presented here in explaining the various constraints, we believe \cdf{} is the orphan afterglow of an off-axis gamma-ray burst. This makes \cdf{} potentially one of the first orphan afterglows ever detected, considering the uncertain nature of other candidates~\citeg{cenko11, marcote19, ho20, sarin21_blt}. The method developed here is ideally suited for detecting orphan afterglows from other X-ray surveys. 

Our inferred parameters, the host galaxy properties, optical upper limits, and inconsistency with the empirical Amati relation imply that \cdf{} is more likely to be the afterglow of a short gamma-ray burst afterglow than a long, implying the most likely progenitor model is that of a binary neutron star coalescence similar to the first multimessenger gravitational-wave detection GW170817~\citep{abbott17_gw170817_detection,abbott17_gw170817_multimessenger}.
This has far-reaching implications. Most notably, at a potential redshift of $z=2.23$, this is likely one of the most distant short gamma-ray burst ever observed (GRB090426, with a redshift of $z=2.61$, was initially classified as a short gamma-ray burst~\citep{antonelli09}, but is now likely considered a long gamma-ray burst~\citep{levesque10}).

Observing neutron star mergers so early in the Universe has implications for our understanding of stellar evolution. The best-fit redshift for \cdf{} places it before the peak of star formation, implying the binary that merged to produce the gamma-ray burst must have formed and merged on a relatively short timescale. 
Population synthesis studies estimate that the rate for short gamma-ray bursts peaks in the redshift range $z = 0.6$--1~\citep{wiggins18}, similar to observations of short gamma-ray bursts with known redshifts~\citep{fong15}.
The existence of \cdf{} and its potential progenitor being the merger of a neutron star binary suggests that the rate of the merger of these systems at high redshifts is not negligible, which has important implications for understanding binary stellar evolution and for detecting gravitational waves from mergers of these objects. 
Furthermore, it may imply that there are significant short delay-time binary neutron star mergers, which may mean that a relatively high fraction of binary mergers occurs in dense stellar environments such as globular clusters, or that common-envelope evolution is more efficient at reducing the orbital separation than previously realised. This also has implications for the offset of short gamma-ray bursts from their host galaxy, particularly at higher redshifts.

A neutron star merger progenitor hypothesis for \cdf{} also has implications for heavy element nucleosynthesis.
The multimessenger observations of GW170817 confirmed that binary neutron star mergers are the production sites of heavy $r$-process elements such as gold, which are difficult to produce in ordinary supernovae. 
However, owing to the relative lack of expected mergers at high redshifts, they are not believed to be the only source~\citep{siegel19}.
If instead there are more neutron star mergers at high redshifts than otherwise expected, this could imply that these systems play a more critical role in the chemical evolution of the Universe.


\section{Acknowledgements}
We are grateful to Carl Knox for the creation of the graphics in Fig.~\ref{fig:lightcurveandcounts}, and Franz Bauer for help with collecting and processing the data. We also thank Geoffrey Ryan for their help in understanding the \afterglowpy{} model. We are grateful to Ilya Mandel, Ryan Shannon, and Matthew Bailes for their valuable comments and discussions.
Computations were performed on the OzStar supercomputer.
N.S is supported by an Australian Government Research Training Program (RTP) Scholarship. P.D.L. is supported through Australian Research Council Future Fellowship FT160100112 and CE170100004. P.D.L and G.A are supported by ARC Discovery Project DP180103155.


\bibliographystyle{apsrev4-1} 
\bibliography{ref}

\end{document}